\documentclass[pra,twocolumn,showpacs]{revtex4-1}
\usepackage{amsmath,amscd,amssymb,color}
\usepackage{graphicx,amsfonts,epsf}
\usepackage{multirow}
\usepackage{epstopdf}
\usepackage{float}
\usepackage{hyperref}
\usepackage{enumerate}
\hypersetup{colorlinks,breaklinks,
linkcolor=blue,urlcolor=blue,
anchorcolor=blue,citecolor=
blue,pdfauthor=Kavita-Dorai, pdftitle=NMR-Relaxation}
\begin{document}
\title{Using a Lindbladian approach to model
decoherence in two coupled\\
nuclear spins via correlated phase-damping and amplitude damping
noise channels} 
\author{Harpreet Singh}
\email{harpreet.singh@tu-dortmund.de}
\affiliation{Department of Physical Sciences, Indian
Institute of Science Education \&
Research Mohali, Sector 81 SAS Nagar,
Manauli PO 140306 Punjab India.}
\affiliation{Fakult{\"a}t Physik,
Technische\\ Universit{\"a}t Dortmund D-44221, 
Dortmund, Germany.}
\author{Arvind}
\email{arvind@iisermohali.ac.in}
\affiliation{Department of Physical Sciences, Indian
Institute of Science Education \&
Research Mohali, Sector 81 SAS Nagar,
Manauli PO 140306 Punjab India.}
\author{Kavita Dorai}
\email{kavita@iisermohali.ac.in}
\affiliation{Department of Physical Sciences, Indian
Institute of Science Education \&
Research Mohali, Sector 81 SAS Nagar,
Manauli PO 140306 Punjab India.}
\begin{abstract}
In this work, we studied the relaxation dynamics of
coherences of different order present in
a system of two coupled nuclear spins.
We used a
previously designed model 
for intrinsic noise present in such systems which considers
the Lindblad master equation for Markovian relaxation.
We experimentally created zero-, single- and double- quantum
coherences in several two-spin systems and performed
a complete state tomography and computed state fidelity.
We experimentally measured the decay
of zero- and double- quantum coherences in these systems.
The experimental data fitted well 
to a model that considers the main
noise channels to be a correlated phase damping channel acting
simultaneously on both spins in conjunction with a generalized
amplitude damping channel acting independently on both spins.
The differential relaxation of multiple-quantum coherences can be
ascribed to the action of a correlated phase damping channel acting 
simultaneously on both the spins.
\end{abstract}
\pacs{03.65.Yz, 76.60.-k, 03.67.a}
\maketitle
\section{Introduction}
Quantum coherence can be associated with a transition between 
the eigenstates of
a quantum system and most spectroscopic signals crucially rely on 
the manipulation, transfer and detection of 
such coherences~\cite{streltsov-rmp-17}. 
In nuclear magnetic resonance (NMR), spin coherence resides in the off-diagonal elements of the density
operator of the system and a system of coupled spin-1/2 nuclei can have
coherences of different orders 
$n$ ($n = 0, 1, 2....$)~\cite{ernst-book-87}.  
NMR is able to directly access only those
off-diagonal elements of the density matrix whose difference in magnetic
quantum number is $\pm 1$ (the single quantum transitions).  The direct
observation of multiple quantum transitions ($\Delta m \ne \pm 1$) is forbidden
by quantum-mechanical selection rules (in the
dipole approximation).  Multiple quantum coherences have found
several useful applications in NMR including spectral simplification,
spin-locking and cross-polarization experiments~\cite{pines-mq}.

The interaction with the environment of a quantum system  causes
loss of coherence and forces the system to relax back towards
a time-invariant equilibrium state.
This limits the time over which coherences live and leads to
poor signal sensitivity~\cite{Schlosshauer-rmp-05}.
In solution NMR the problem is
exaggerated when dealing with larger spin systems such as those encountered in
proteins, where slower rotational tumbling of the molecules leads to faster
rates of relaxation, and consequently larger 
losses in signal~\cite{cavanagh-book}.
Coherence preservation is hence of supreme importance in NMR experiments and
several schemes have been designed to suppress spin
relaxation including using
longlived two-spin order states which have lifetimes much longer than T$_1$,
termed singlet states~\cite{singlet-2spin-review,jmr-singlet-bangbang-tsm}.
Links between NMR coherence orders and decoherence have
been recently investigated~\cite{brazil-coh-order}.
Several 
NMR techniques have benefited from cross-fertilization of
ideas from other fields of research such as quantum information processing.
For instance, several methods that suppress spin relaxation such as
optimal control 
theory~\cite{tosner-jmr-09}
and dynamical decoupling~\cite{suter-dd,singh-pra-14}, 
have all drawn on insights from
their initial application to general problems of 
quantum decoherence, algorithm implementation~\cite{pramana1} and
quantum entanglement~\cite{pramana2}.  
Recently, certain special types of correlation
functions termed out-of-time-order correlations (OTOC) have
been used to characterize the delocalization of quantum
information and have been linked to multiple-quantum
coherences~\cite{otoc-review,otoc-cory,otoc-nmr-jfdu,otoc-tsm}.
Optimal control techniques have also been used to control
coupled heteronuclear spin dynamics in the presence of
general relaxation mechanisms and to explore how closely a
quantum system can be steered to a target
state~\cite{luy1,luy2,glaser-ccr-grape,glaser-t1-t2-pra,glaser-pnas-relaxation}.
Synthesizer noise can lead to severe dephasing effects 
akin to a decohering environment and
new methods have been recently proposed to eliminate such noise using
two single-spin systems in opposite static 
magnetic fields~\cite{qtmengg1,qtmengg2}.
Transverse relaxation times in systems of coupled spins
have been accurately measured and the noise profiles of
multi-spin coherences and their scaling 
with respect to
coherence order has been studied~\cite{rangeet,tsm-noise}.
In order to devise techniques to 
obviate the deleterious effects of 
spin relaxation, one first needs to gain a
deeper understanding of the mechanisms underlying this complex
phenomenon.
Molecules in a liquid freely tumble and undergo stochastic Brownian motion
which is the main source of NMR spin relaxation, where the spin lattice
degrees of freedom include all the molecular rotational and translational
motions.  The semi-classical Redfield approach is typically used to describe
NMR spin relaxation which uses the density matrix formalism and second order
perturbation theory; the noisy spin environment is treated classically by a
spin lattice model while the spins are treated as quantum mechanical objects
and a weak system-environment coupling is 
assumed~\cite{redfield}.  The bath
correlations decay much more rapidly than the evolution of the spins and the
Markovian approximation remains valid.  
For two coupled spins 1/2, the major relaxation mechanisms in NMR are the
dipole-dipole (DD) relaxation and the relaxation arising from the chemical
shift anisotropy (CSA) of each spin.  In general, interference terms between
the DD and CSA relaxation mechanisms can give rise to another mechanism for
relaxation termed as cross-correlated spin relaxation~\cite{anil-prog,kd-book}.
Extensions of Bloch-Redfield relaxation theory have developed a unified
picture by including contributions from dipolar coupling between
remote spins~\cite{jeener,goldman} and 
by considering a two-state Markov noise process
which includes lattice fluctuations and chemical exchange
dynamics~\cite{abergel}. 
The most general form for the nonunitary evolution of
the density operator of an open quantum system can be described by a master
equation~\cite{lindblad}.  
In the master
equation approach,  both the environment 
and the spins are assumed to be quantum
mechanical in character.  
The Redfield approach is a ``bottom-up'' approach
which begins with the allowed degrees of freedom and the relaxation
mechanisms which are specific to the system under consideration and
then builds a model from them.
The master equation approach on the other hand, is a ``top-down'' approach which
begins by considering all possible allowed relaxation processes and
then concludes from the data which are the noise channels that are dominant.
The insights gained from the Redfield and the master equation
methods are complementary in character, and using a combined approach can
help build a complete picture of coupled spin relaxation.

In the master equation formalism, the NMR longitudinal T$_1$ and transverse
T$_2$  relaxation processes are described by two different noise channels,
namely the amplitude damping and the phase damping channel, 
respectively~\cite{redfield-master-pra}.  The
effect of the phase damping channel on a single spin is to nullify the
coherences stored in the off-diagonal elements of the spin density matrix.  The
generalized amplitude damping channel leads to energy loss through dissipative
interactions between the spin and the lattice at finite temperatures, where the
spin in the excited state decays to its ground state.  
The Lindblad operators
were delineated for a system of two coupled spin-1/2 nuclei
by measuring the density operator at multiple 
time points~\cite{cory-lindblad-pra}.
The
phase damping, amplitude
damping and depolarizing noise channels have been implemented 
in NMR using two and
three heteronuclear coupled spins~\cite{long-qtm-channel}.

It has long been known in NMR that the relaxation of multiple quantum
transitions contains useful information about correlated fluctuations occurring
at different nuclear sites as well about molecular
motions~\cite{ernst-book-87}.  In contrast to single quantum
experiments on coupled spins, the relaxation dispersion profiles of
multiple-quantum relaxation rates are sensitive to the chemical environment of
the involved nuclei and can hence be used to gain insights about millisecond
time-scale dynamics in large 
biomolecules~\cite{tq-cpmg-diffusion-kay}.  Multiple
quantum relaxation has been used to probe protein-ligand interactions,
conformational exchange processes and side-chain motions in
proteins~\cite{methyl-mq-trosy}.

A spin system consisting of coupled spins of the same nuclear species is termed
a homonuclear system while a coupled spin system consisting of different
nuclear species is called a heteronuclear system.
In this work, we focus on studying the relaxation dynamics 
of quantum coherences in
homonuclear systems of coupled spin-1/2 nuclei.
On the other hand, in a heteronuclear coupled two-spin system,
the noise was fitted using several
noise models and it was shown that such systems  
can be treated
as being acted upon by independent noise channels~\cite{childs-pra-01}.  
We use the Lindblad master equation for Markovian relaxation to set up
and analyze the relaxation of coherences of different order, namely zero-,
single-, and double-quantum coherences.
We first experimentally prepared states with different orders of quantum
coherences, tomographed the state, and computed state fidelity.  We then
allowed the state to decay and experimentally measured the decay rates of
different quantum coherences.  The experimentally determined evolution of the
density matrices for the states prepared as pure double-, single-, or
zero-quantum coherence were obtained via a generalized master equation
formalism.  We modeled the inherent noise in the system by assuming that a
correlated phase damping quantum channel acts on both spins and that a
generalized amplitude damping quantum channel acts independently on each spin.
We obtained good fits of the theoretical model to the experimental data within
reasonable experimental errors.  It has been shown that cross-correlated spin
relaxation terms arising 
both from auto-correlation spectral densities for
dipolar relaxation as well as
from
a ``remote'' CSA-CSA cross-correlation
mechanism, contribute differentially to the relaxation of zero- and
double-quantum coherences in a coupled two-spin
system~\cite{pkumar}.  We conjecture that the Redfield description
of CSA-CSA cross-correlated spin relaxation is analogous to the correlated phase
damping channel in the generalized master equation description of relaxation.
The distinctly different relaxation dynamics of multiple-quantum coherences
in homonuclear coupled spin systems as opposed to heteronuclear systems is
clearly evident from our analysis. The relaxation behavior of homonuclear
coupled spin-1/2 nuclei can be explained on the basis of a correlated phase
damping noise channel acting on both spins and 
ties in well with the standard Redfield method of studying spin relaxation.
\section{Modeling intrinsic noise in NMR}
\label{dynamics}
The operator-sum representation is typically used to describe
quantum decoherence~\cite{nielsen-book-02}.
A noisy channel acting on an input density matrix $\rho$ is given by a
completely positive trace preserving map 
\begin{equation}
\mathcal{E} (\rho) = \Sigma_i E_i \rho E_i^{\dagger}
\end{equation}
where $E_i$ are Kraus operators describing a noisy channel in the sum-operator
approach and $\Sigma_k E_k E_k^{\dagger} = 1$ ensures that unit trace is
preserved.
The final noisy state is~\cite{correlated-noise,maniscalco-correlated}
\begin{equation}
\mathcal{E}(\rho) = 
(1-\mu) \Sigma_{i,j} E_{i,j} \rho E_{i,j}^{\dagger} +
\mu \Sigma_{k} E_{k,k} \rho E_{k,k}^{\dagger} 
\end{equation}
where $\mu$ is probability for noise to be correlated and
$(1-\mu)$ is the probability for uncorrelated noise.

For a special class of noisy channels where the Markovian
approximation is valid, 
one can write the master equation governing decoherence in
a Lindblad form~\cite{lindblad,jung-pra-08,daffer,njp2010}
\begin{equation}
\frac{\partial \rho}{\partial t} 
= \sum_{i,\alpha}
\left[
L_{i,\alpha} \rho L_{i,\alpha}^{\dagger}
- \frac{1}{2} \{ L^{\dagger}_{i,\alpha} 
L_{i,\alpha},\rho
\}
\right]
\label{mastereqn} 
\end{equation}
where $L_{i,\alpha} \equiv \sqrt{\kappa_{i,\alpha}} 
\sigma^{(i)}_{\alpha}$ is the Lindblad operator
and $\sigma^{(i)}_{\alpha}$
is the Pauli operator of the $i$th spin
($\alpha=x,y,z$) and the constant
$\kappa_{i,\alpha}$ has the units of inverse  
time.
It has been proved
that a linear operator on a finite $N$-dimensional
Hilbert space is the generator of a completely positive dynamical
semigroup~\cite{gorini} and hence the Lindbladian is the generator of the 
semigroup which governs the dissipation of the density operator.

In the language of the master equation approach to decoherence, the
relaxation of an NMR spin tumbling isotropically in a solution can be described
by two noise channels: a phase damping channel and an amplitude damping
channel~\cite{Paula-prl-13,auccaise-prl-11,singh-epl-17,silva-prl-16}.  
Due to molecular tumbling the average magnetic field experienced by a spin over
time is the same but it varies across the sample at a particular time which
causes identical spins to slowly lose phase coherence, which is the process of
phase damping.  The Kraus superoperator for the phase damping (PD) channel
acting on a single spin density operator ($\rho$) can be written as
\begin{equation}
 {\cal E}^{\rm PD}(\rho)=\left(\begin{array}{cc}
 \rho_{00} & \rho_{01}e^{-\gamma t} \\
 e^{-\gamma t}\rho_{10} &\rho_{11} \\
\end{array}
\right)
\end{equation}
for a damping rate $\gamma$.
The generator of PD channel on a single spin can be written as
\begin{equation}
 {\cal Z}^{\rm PD}(\rho)=-\gamma\left(\begin{array}{cc}
 0 & \rho_{01} \\
 \rho_{10} &0 \\
\end{array}
\right)
\end{equation}
Ordering the matrix elements of $\rho$ in the vector\\ $(\rho_{00}, \rho_{01},
\rho_{10}, \rho_{11})^T$, we have
\begin{equation}
{\cal Z}^{\rm PD}=\left(\begin{array}{cccc}                     
 0 & 0       & 0       & 0 \\
 0 & -\gamma & 0       & 0 \\
 0 & 0       & -\gamma & 0 \\
0 & 0 & 0 & 0
\end{array}\right)
\end{equation}
The generalized amplitude damping channel (GAD) for a single spin
models the process where the spin exchanges energy
with a reservoir at some fixed temperature 
\begin{equation}  
{\cal E}^{\rm GAD}(\rho) =\left(\begin{array}{cccc}   
   {k_1 \rho_{00} + k_2 \rho_{11}}&{e^{-\Gamma t/2} \rho_{01}} \cr
    {e^{-\Gamma t/2} \rho_{10}}&{k_3 \rho_{00} + k_4 \rho_{11}}
\end{array}\right)
\end{equation}
where $k_2 \equiv (1-\bar n)(1-e^{-\Gamma t})$, $k_3 \equiv \bar
n(1-e^{-\Gamma t})$, $k_1 \equiv 1-k_3$, and $k_4 \equiv 1-k_2$, 
$\Gamma$ is the damping rate and $\bar n$ is a temperature parameter 
\begin{equation}
\log{1-\bar n \over \bar n} = {\Delta E \over k_B T  }
\end{equation}
with $\Delta E$ being the energy level difference between 
the ground and excited states of
the system.

An ensemble of NMR spins in thermal equilibrium at room temperature 
has a Boltzmann distribution of spin populations and in this
high-temperature limit,
the generator of the GAD channel  is given by
\begin{equation}
{\cal Z}^{\rm GAD^{\infty}} = -\Gamma \left(\begin{array}{cccc}  
\frac{1}{2}   & 0           & 0           &-\frac{1}{{2}} \cr
0       & {1 \over 2} & 0           & 0   \cr
 0       & 0           & {1 \over 2} & 0   \cr
-\frac{1}{{2}}& 0           & 0           & \frac{1}{{2}} 
                 \end{array}\right)
\end{equation}

The simplest extension of these single-spin decoherence processes to a two-spin
system is to consider a phase damping and an generalized amplitude damping
channel acting independently on each spin.  However, since the spin systems we
have studied are homonuclear, with two proton nuclear species having slightly
different Larmor resonance frequencies, we hypothesize 
that each spin decoheres under the concerted action
of a phase-damping channel which is correlated with that of the other spin.
Hence, the naturally occurring decoherence for this system can be modeled as a
correlated dephasing channel acting on both spins and a generalized amplitude
damping channel acting independently on each spin~\cite{childs-pra-01}. 
The generator of the correlated phase damping channel acting on both
spins is given by:
\begin{eqnarray}
{\cal Z}^{\rm CPD}={\rm
diag}&[&0,-\gamma_2,-\gamma_1,-(\gamma_1+\gamma_2+\gamma_3), \nonumber\\
     && -\gamma_2,0,-(\gamma_1+\gamma_2-\gamma_3),-\gamma_1, \nonumber\\
     && -\gamma_1,-(\gamma_1+\gamma_2-\gamma_3),0,-\gamma_2, \nonumber\\
     && -(\gamma_1+\gamma_2+\gamma_3),-\gamma_1,-\gamma_2,0]
\,,
\end{eqnarray}
where $\gamma_1$ and $\gamma_2$ are  decay rates for independent
phase damping on spins $1$ and $2$, and $\gamma_3$ can be interpreted as a
rate for correlated phase damping.  
The full generator to describe two-spin decoherence has the form
\begin{equation}
\label{pdampmodel}
  {\cal Z}=   {\cal Z}^{\rm CPD}
           + {\cal Z}^{\rm GAD^{\infty}}_1 + {\cal Z}^{\rm GAD^{\infty}}_2
\,,
\end{equation}

Under the action of the full decoherence generator ${\cal Z}$, 
the state $\rho$
decoheres to:
\begin{equation}
 {\cal E}^{{\rm 2spin}}(\rho)=\left(
\begin{array}{cccc}
\alpha_1 & \beta_1 &\beta_2  & \beta_3  \\
 \beta_1  & \alpha_2& \beta_4  & \beta_5 \\
 \beta_2 &\beta_4 & \alpha_3 & \beta_6 \\
 \beta_3  & \beta_5 & \beta_6 & \alpha_4 \\
\end{array}
\right)
\label{superopdecay}
\end{equation}
The parameters $\alpha_i, \beta_i$ can be written in
terms of the decay rates $\gamma_i, i=1,2,3$ 
and $\Gamma_i, i=1,2$
of the correlated PD channel and the 
independent GAD channels, respectively. In the next section, 
we will proceed towards the explicit
calculation of the superoperator ${\cal E}^{{\rm 2spin}}(\rho)$ for different
input states.
\section{Results and Discussion}
\label{system}
\subsection{System Details}
In high field NMR, the Zeeman interaction causes a splitting of the energy
levels according to the field direction and the difference between magnetic
quantum numbers $\Delta m_{rs}= m_r-m_s$ defines the 
order of the coherence~\cite{ernst-book-87}.  
For two-spin systems, if
$\Delta m_{rs}=0$ the coherence is a zero quantum (ZQ) coherence, if $\Delta
m_{rs}=\pm1$ the coherence is a single quantum (SQ) coherence, and if $\Delta
m_{rs}=\pm2$ the coherence is a double quantum (DQ) coherence.  

\begin{table}[htb]
\label{parameter-table}
\caption{NMR parameters of homonuclear two-spin systems used in this
study.}
\vspace*{12pt}
\centering
\renewcommand{\arraystretch}{1.5}
\begin{tabular}{llll}
\hline
Molecule &
$(\nu_1, \nu_2)$ (Hz)&$\Delta \nu$ (Hz) &
$J_{12}$ (Hz)~~~\\
\hline
BTC acid& ($4602.4$, $4287.0$)&315.4 & 4.2~~~\\
Cytosine & $(4407.7,3490.8)$& 916.9& $7.1$~~~\\
Coumarin&($4734.0$, $3807.9$)&926.1 & 9.5~~~\\
\hline
\end{tabular}
\end{table}

\begin{figure}[h]
\includegraphics[angle=0,scale=1.0]{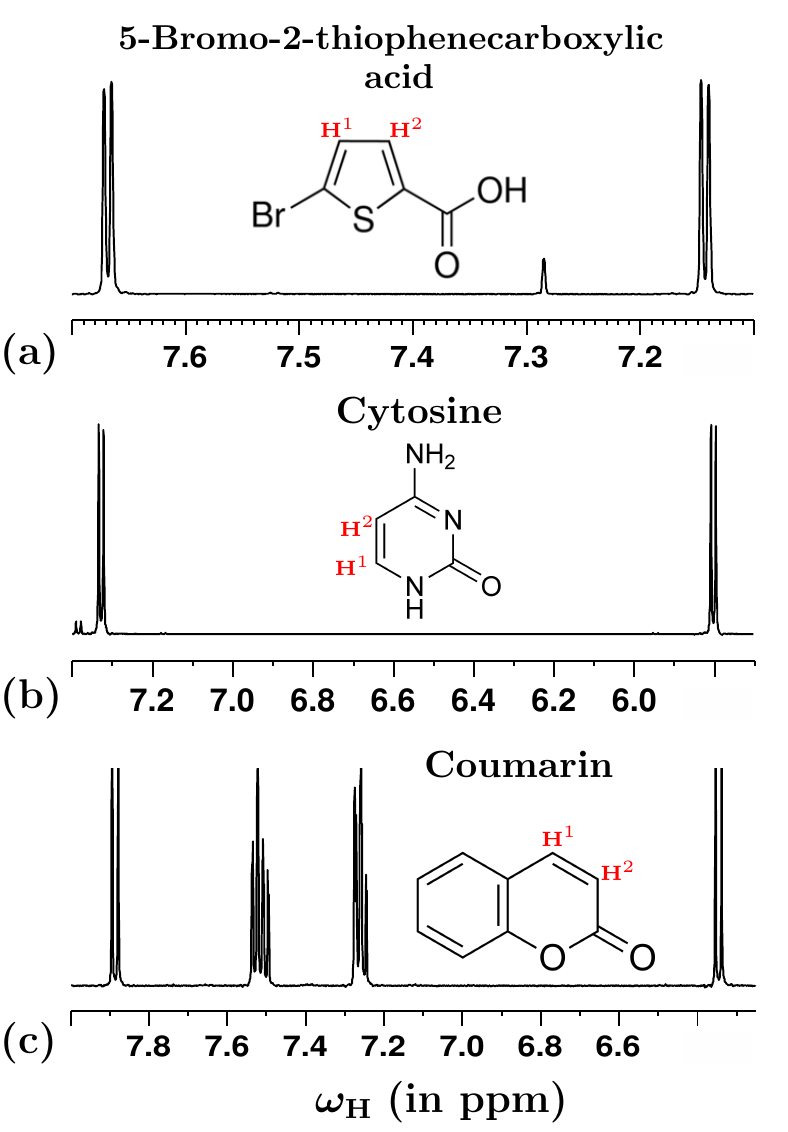}
\caption{NMR spectra obtained after a $\pi/2$
readout pulse on the thermal equilibrium state of
(a) 5-bromo-2-thiophenecarboxylic (BTC) acid, (b) Cytosine
and (c) Coumarin.
}
\label{molecule}
\end{figure}

The Hamiltonian of a weakly-coupled two-spin system in a frame rotating at
$\omega_{rf}$ in a static magnetic field $B_0$ is given by \begin{equation} H =
-(\omega_1-\omega_{rf}) I_{1z} - (\omega_2-\omega_{rf}) I_{2z} + 2\pi J_{12}
I_{1z}I_{2z} \end{equation} where $I_{iz}$ is the $z$th component of the spin
angular momentum operator,  the first two terms in the Hamiltonian denote the
Zeeman interaction between each spin and the static magnetic field $B_0$, and
the last term represents the spin-spin interaction  with $J_{ij}$ being the
scalar coupling constant.  We used the $^1\rm{H}$ spins of
5-bromo-2-thiophene-\\carboxylic (BTC) acid,
cytosine, and coumarin as model
homonuclear two-spin systems.  The molecular structure of these two-spin
systems and the NMR spectra of the spins at thermal equilibrium are shown in
Figs.~\ref{molecule}(a), (b), and (c), respectively.  The experiments were
performed at an ambient temperature of 298 K on a Bruker Avance III 600 MHz NMR
spectrometer equipped with a QXI probe. 
\subsection{State initialization schemes}
We initialize our system into a ``pseudopure'' state, wherein all the
energy levels except one, are uniformly populated. Such special
quantum states
have interesting properties and have recently found several applications
in the area of quantum information processing~\cite{nielsen-book-02}.
While standard schemes for pseudopure state preparation
involve a large number of experiments and lead to reduced signal, recently
a few schemes have been proposed that use only one ancilla spin and fewer
number of experiments~\cite{china1,china2}.
The relaxation
behavior of two-spin pseudopure states have
been investigated and it was noted that cross-correlated spin relaxation plays
an important role in accelerating or retarding the lifetimes of such
states~\cite{jmr-anil-ccr-ppure}.  
The relaxation of pseudopure states in an oriented spin-3/2
system has been described using Redfield theory and reduced spectral
densities~\cite{jmr-quadrupolar-nmr-relaxation}.

The two-spin
equilibrium density matrix (in the high temperature 
and high field approximations)
is in a highly mixed state given by:
\begin{eqnarray}
\rho_{eq}&=&\tfrac{1} {4}(I+\epsilon \ \Delta \rho_{eq})
\nonumber \\
\Delta\rho_{{\rm eq}} &\propto& \sum_{i=1}^{2} I_{iz}
\end{eqnarray}
with a thermal polarization $\epsilon \sim 10^{-5}$, $I$
being an $4 \times 4$ identity operator and 
$\Delta \rho_{{\rm eq}}$ being the deviation part of
the density matrix.

\begin{figure}[h]
\includegraphics[angle=0,scale=1]{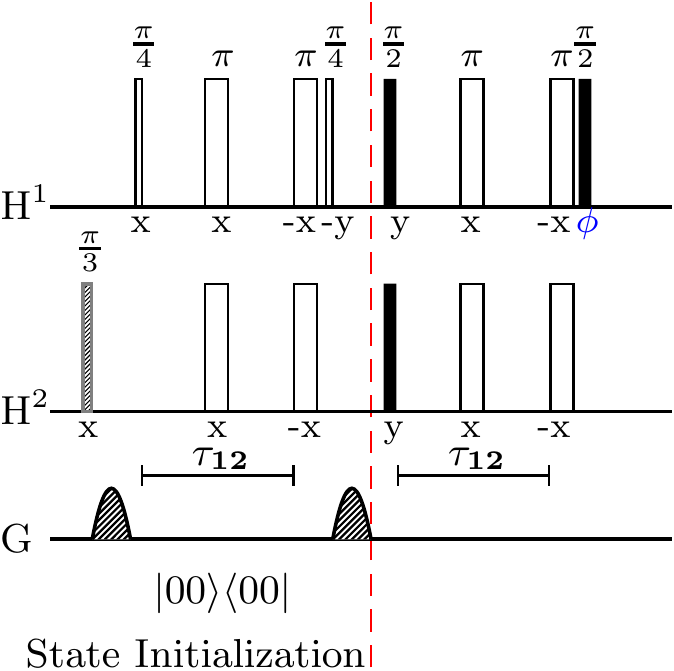}
\caption{Pulse sequence for the preparation of 
$\frac{1}{\sqrt{2}}(\vert00\rangle+\vert11\rangle)$ and
$\frac{1}{\sqrt{2}}(\vert01\rangle+\vert10\rangle)$ states 
from thermal equilibrium. The sequence of pulses before
the vertical dashed red line achieves state 
initialization into the $\vert00\rangle$ pseudopure state. 
Filled  
and unfilled rectangles represent $\frac{\pi}{2}$ and $\pi$
pulses respectively, 
while all other rf pulses are labeled with their
respective flip angles.  The phase of the rf pulse is written
below each pulse, with
the phase $\phi$ kept along x(-x), depending on the desired coherence
order; $\tau_{12}$ denotes a delay fixed at $1/2J_{12}$. 
}
\label{pulse_sequence}
\end{figure}

We use the notation $\vert 0 \rangle$ to denote the eigenstate of
a spin-1/2 particle in the ground state (spin ``up'')
and $\vert 1 \rangle$ to denote
the eigenstate of the excited state (spin ``down'').
The two-spin systems were initialized into the $\vert 00\rangle$
pseudopure state using the spatial averaging
technique~\cite{cory-physicad},
with the density operator given by
\begin{equation}
\rho_{00}=\frac{1-\epsilon}{4}I 
+ \epsilon \vert 00\rangle\langle00 \vert
\end{equation} 
The pulse sequence for the preparation of  $\vert 00\rangle$
from thermal state is shown in the first part of Fig.\ref{pulse_sequence}.
The pulse propagators for selective excitation were
constructed using the GRAPE algorithm~\cite{tosner-jmr-09} to design
the amplitude and phase modulated RF profiles.  
Numerically generated GRAPE pulse profiles were optimized to be robust
against RF inhomogeneity and had an average fidelity of 
$ \ge 0.995$. Selective excitation was typically
achieved with pulses of duration 10 ms for 
BTC acid and 1 ms for both coumarin 
and cytosine molecules.
\subsection{Final density matrix reconstruction}
We interrogate our final density matrix via a useful technique
called quantum state tomography, which uses a set of
measurements of the expectation values of spin angular
momentum operators, to independently quantify all the
real and imaginary elements of the density matrix.
One can hence specifically follow the relaxation rates
of different elements of the 
density matrix~\cite{long-qst}.
All experimental density matrices were reconstructed using a
reduced tomographic protocol~\cite{leskowitz-pra-04,singh-pla-16}, with
the set of operations given by $\{II, IX, IY, XX\}$ being
sufficient to determine all 15 variables for the two-spin
system.  Here $I$ is the identity (do-nothing operation) and
$X (Y)$ denotes a single spin operator that can be
implemented by applying a spin-selective
$\pi/2$ pulse on the corresponding spin. 
\subsection{Measuring state fidelity}
The fidelity is an estimate of the ``closeness'' between two
pure states or between two density matrices.
The fidelity of an experimental
density matrix was computed by 
measuring the
projection between the
theoretically expected and experimentally
measured states using the Jozsa and Uhlmann
fidelity measure~\cite{jozsa-fidelity,uhlmann-fidelity}:
\begin{equation}
F =
\left(Tr \left( \sqrt{
\sqrt{\rho_{\rm theory}}
\rho_{\rm expt} \sqrt{\rho_{\rm theory}}
}
\right)\right)^2
\label{fidelity}
\end{equation}
where $\rho_{\rm theory}$ and $\rho_{\rm expt}$ denote the
theoretical and experimental density matrices, respectively. 
In our experiments, we use fidelity as a measure to evaluate how 
well our experimental schemes were able to achieve the theoretically
expected final density matrices.
\subsection{Experimental creation of multiple-quantum coherences}
The desired order of multiple-quantum coherence i.e. zero- or double- was
prepared using the latter part (after red dashed line) of the pulse sequence
given in Fig.\ref{pulse_sequence}.  A non-selective $\frac{\pi}{2}$ pulse was
applied along the $y$-axis, which rotates both spins onto the $x$-axis,
followed by a delay of $\tau=\frac{1}{2J_{12}}$ (along with refocusing pulses
being applied at the center and at the end of the delay).  A GRAPE-optimized
spin-selective $\frac{\pi}{2}$ pulse is applied along -x(x) axis in order to
prepare either the zero- quantum coherence
$\frac{1}{\sqrt{2}}(\vert01\rangle+\vert10\rangle)$ or the double-quantum
coherence $\frac{1}{\sqrt{2}}(\vert00\rangle+\vert11\rangle)$.  We were able to
achieve final state fidelities of $ \approx 0.99$ for all the three homonuclear
spin systems studied.

\begin{figure}[h]
\includegraphics[angle=0,scale=1]{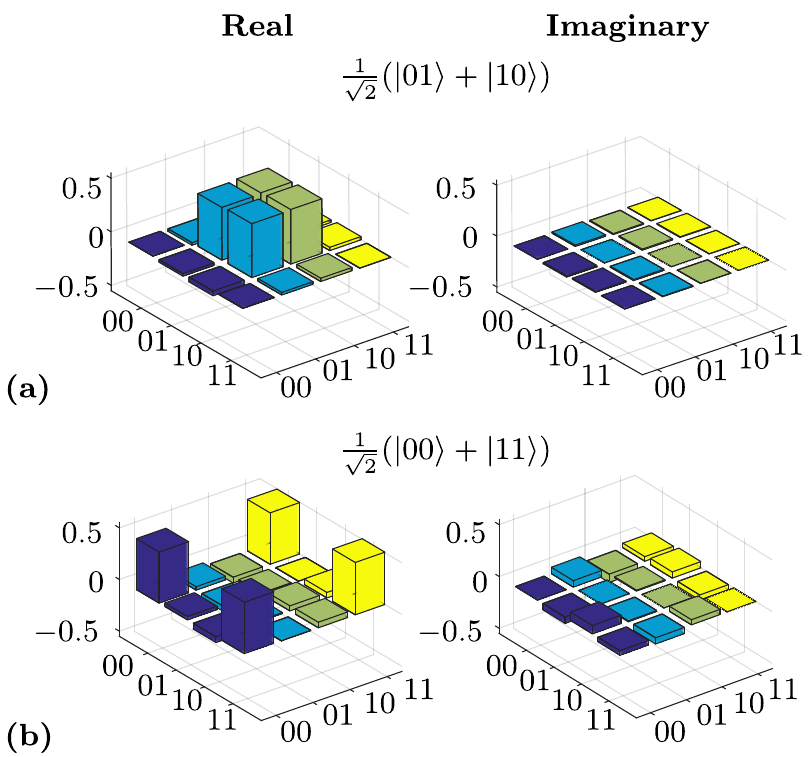}
\caption{
The real (left) and imaginary (right) parts
of the experimentally tomographed 
density matrix of the BTC acid molecule in the
(a) $\frac{1}{\sqrt{2}}(\vert01\rangle+\vert10\rangle)$
state, with a fidelity of 0.98 and in the (b) 
$\frac{1}{\sqrt{2}}(\vert00\rangle+\vert11\rangle)$ state,
with a fidelity of 0.99. The rows and columns encode the
computational basis in binary order from 
$\vert00\rangle$ to $\vert11\rangle$.}
\label{tomo_sample1}
\end{figure}

\begin{figure}
\includegraphics[angle=0,scale=1]{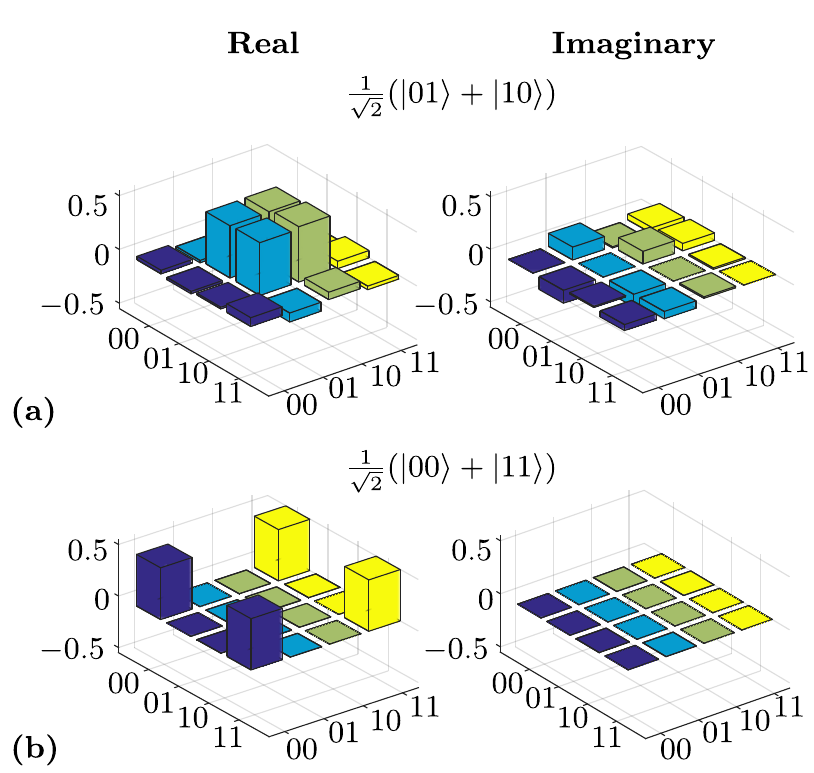}
\caption{The real (left) and imaginary (right) parts
of the experimentally tomographed density matrix of the
coumarin molecule in the
(a) $\frac{1}{\sqrt{2}}(\vert01\rangle+\vert10\rangle)$
state, with a fidelity of 0.98 and
in the (b) $\frac{1}{\sqrt{2}}(\vert00\rangle+\vert11\rangle)$ state,
with a fidelity of 0.99. The rows and columns encode the
computational basis in binary order from 
$\vert00\rangle$ to $\vert11\rangle$.}
\label{tomo_sample2}
\end{figure}

\begin{figure}
\includegraphics[angle=0,scale=1]{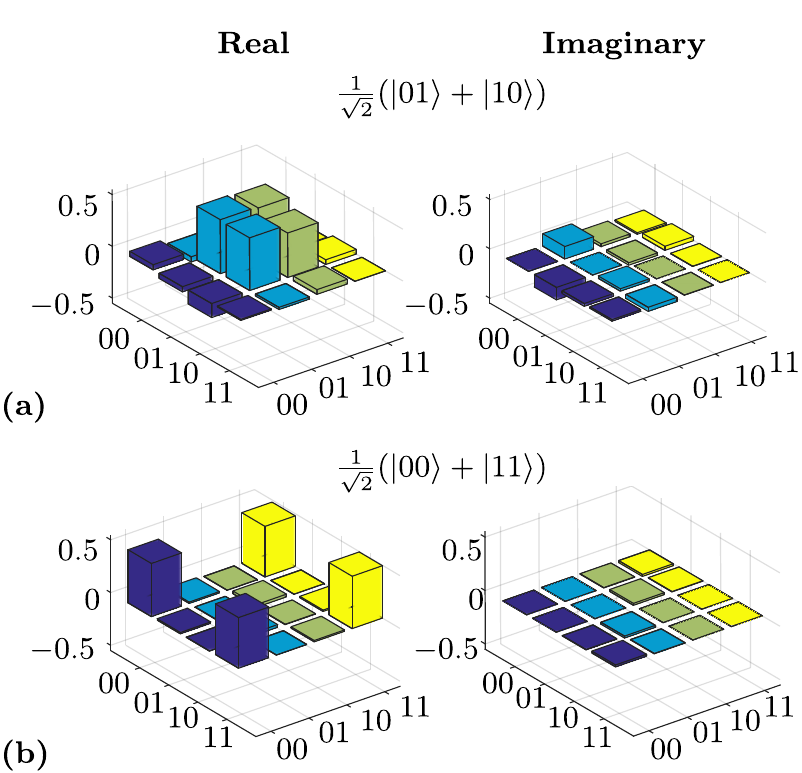}
\caption{
The real (left) and imaginary (right) parts
of the experimentally tomographed 
density matrix of the cytosine molecule in the
(a) $\frac{1}{\sqrt{2}}(\vert01\rangle+\vert10\rangle)$
state, with a fidelity of 0.98 and in
the (b) $\frac{1}{\sqrt{2}}(\vert00\rangle+\vert11\rangle)$ state,
with a fidelity of 0.99. The rows and columns encode the
computational basis in binary order from 
$\vert00\rangle$ to $\vert11\rangle$.}
\label{tomo_sample3}
\end{figure}

Figs.~\ref{tomo_sample1}-\ref{tomo_sample3} (a)-(b)
depict the real (left panel) and imaginary (right panel) parts of 
the experimentally reconstructed 
density matrices of the
zero-quantum coherence 
($\frac{1}{\sqrt{2}}(\vert01\rangle+\vert10\rangle)$ state)
and the double-quantum coherence
($\frac{1}{\sqrt{2}}(\vert00\rangle+\vert11\rangle)$ state), respectively,
for the BTC acid, coumarin and cytosine molecules.
Computed state fidelities were
$0.982\pm 0.011$,
$0.983\pm 0.017$,
and $0.983\pm 0.015$ 
for the zero-quantum coherence and
$0.994\pm 0.013$,
$0.991\pm 0.015$,
and $0.979\pm 0.016$ 
for the double-quantum coherence. 

States with single-quantum  coherences, 
$\frac{1}{\sqrt{2}}(\vert00\rangle + \vert10\rangle)$
or $\frac{1}{\sqrt{2}}(\vert00\rangle + \vert01\rangle)$ 
were prepared by applying a $\frac{\pi}{2}$
selective pulse along the $y$-axis on the first (second) spin,
respectively, with computed state fidelities of
$\approx 0.99$.
\subsection{Decay of populations and single-quantum coherences}
Spin-lattice relaxation rates $\Gamma=\rm 1/T_1$ was measured using the
standard  $ 180^{\circ}_y - \tau - 90^{\circ}_x$ inversion recovery pulse
sequence.  The spin-spin relaxation rate $\gamma=\rm 1/T_2$ which is the rate
at which a single-quantum coherence decays, was experimentally measured by
first rotating the magnetization of the spin into the transverse plane by a
$\frac{\pi}{2}$ rf pulse followed by a delay and fitting the resulting
magnetization decay.

The decay of single-quantum coherences with time is shown in
Figs.~\ref{btp_decay}-\ref{coumarine_decay}(a)-(b), for the two-spin
homonuclear systems of BTC acid, cytosine and coumarin, respectively.
The experimentally measured values of spin-lattice relaxation rates $\Gamma_1$
and $\Gamma_2$ in these systems were obtained to be: $\Gamma_1={\rm
1/T_1^{1H}}=0.264\pm0.004$ s$^{-1}$ and $\Gamma_2={\rm
1/T_1^{2H}}=0.255\pm0.003$ s$^{-1}$ for the BTC molecule,  $\Gamma_1=\rm
1/T_1^{1H}=0.153\pm 0.002$ s$^{-1}$ and $\Gamma_2=\rm
T_1^{2H}=0.152\pm0.014$ s$^{-1}$ for the cytosine molecule,  and $\Gamma_1={\rm
1/T_1^{1H}}=0.210\pm 0.004$ s$^{-1}$ and $\Gamma_2={\rm
1/T_1^{2H}}=0.135\pm0.002$ s$^{-1}$ for the coumarin molecule. 
The single-quantum coherence decay rates turned out to be $\gamma_1={\rm
1/T_2^{1H}}=3.741\pm0.242$ s$^{-1}$ and $\gamma_2=\rm
{1/T_2^{2H}}=3.048\pm0.376$ s$^{-1}$ for spin 1 and spin 2, respectively in the
BTC molecule.  In the cytosine molecule, the single-quantum coherence decay
rates were obtained as $\gamma_1=\rm 1/T_2^{1H}=1.618\pm 0.080$ s$^{-1}$ and
$\gamma_2=\rm 1/T_2^{2H}=1.891\pm 0.096$ s$^{-1}$ for spin 1 and spin 2,
respectively.  In the coumarin molecule, the single-quantum coherence decay
rates were obtained as $\gamma_1=\rm  1/T_2^{1H}= 6.813 \pm 0.356$ s$^{-1}$ and
${\gamma_2=\rm 1/T_2^{2H}}=6.761\pm0.286$ s$^{-1}$ for spin 1 and spin 2,
respectively.
\subsection{Decay of multiple-quantum coherences}
\label{exp_decay}
The zero (double)-quantum coherences relaxation rates were experimentally
measured by first preparing either the
$\frac{1}{\sqrt{2}}(\vert01\rangle+\vert10\rangle)$ or the
($\frac{1}{\sqrt{2}}(\vert00\rangle+\vert11\rangle)$) from the thermal 
state, followed by a delay and
then rotating the magnetization of the the first spin by a $\frac{\pi}{2}$
rf pulse and finally, a measurement of the magnetization of the
second spin.  The resulting magnetization decay of the second spin was fitted
to the noise model to obtain an estimate of the multiple-quantum relaxation
rates.

When the initial state is
$\frac{1}{\sqrt{2}}(\vert01\rangle+\vert10\rangle)$
i.e. 
a zero-quantum coherence, the parameters $\alpha_i, \beta_i$
in Eqn.~\ref{superopdecay} are given in terms of the decay rates of the
$\gamma$ of the uncorrelated and correlated PD channels and 
decay rates $\Gamma$ of the independent GAD channels by: 
\begin{eqnarray} 
 \alpha_1 &=& \frac{1}{4}(1-e^{-t (\Gamma_1+\Gamma_2}))\nonumber \\
\alpha_2 &=& \frac{1}{4}(1+e^{-t (\Gamma_1+\Gamma_2}))\nonumber \\
\alpha_3 &=& \frac{1}{4}(1+e^{-t (\Gamma_1+\Gamma_2}))\nonumber \\
 \alpha_4 &=& \frac{1}{4}(1-e^{-t (\Gamma_1+\Gamma_2}))\nonumber \\
\beta_1&=& \beta_2 = \beta_3 = 0 \nonumber \\
\beta_4&=& \frac{1}{2}(e^{-t(\gamma_1+\gamma_2-\gamma_3+\frac{\Gamma_1}{2}+\frac{\Gamma_1}{2})}) \nonumber \\
\beta_5&=& \beta_6 = 0 
\end{eqnarray}

When the initial state is
$\frac{1}{\sqrt{2}}(\vert00\rangle+\vert11\rangle)$ i.e a 
double-quantum coherence, the parameters $\alpha_i, \beta_i$ 
in Eqn.~\ref{superopdecay} are given by: 
\begin{eqnarray} 
 \alpha_1 &=& \frac{1}{4}(1+e^{-t (\Gamma_1+\Gamma_2}))\nonumber \\
\alpha_2 &=& \frac{1}{4}(1-e^{-t (\Gamma_1+\Gamma_2}))\nonumber \\
\alpha_3 &=& \frac{1}{4}(1-e^{-t (\Gamma_1+\Gamma_2}))\nonumber \\
 \alpha_4 &=& \frac{1}{4}(1+e^{-t (\Gamma_1+\Gamma_2}))\nonumber \\
\beta_1&=& \beta_2 = 0 \nonumber \\
\beta_3&=& \frac{1}{2}(e^{-t(\gamma_1+\gamma_2+\gamma_3+\frac{\Gamma_1}{2}+\frac{\Gamma_1}{2})}) \nonumber \\
\beta_4 &=& \beta_5 = \beta_6 = 0 
\end{eqnarray}

\begin{figure}
\includegraphics[angle=0,scale=1.0]{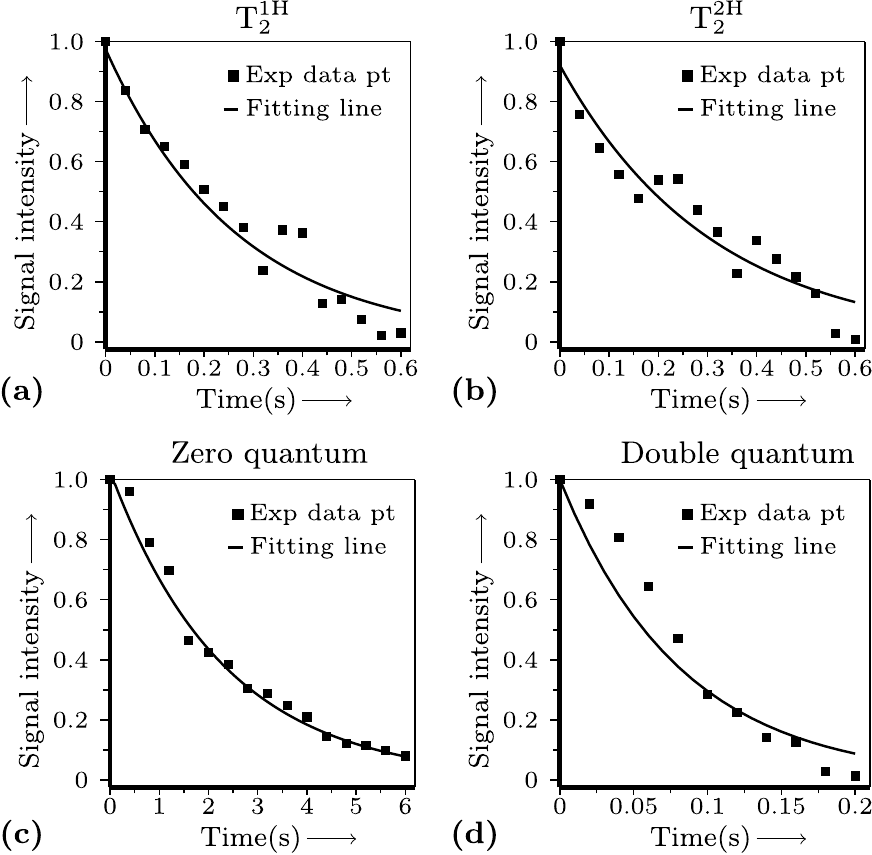}
\caption{Decay of signal intensity with time 
of the (a)single-quantum coherence of spin 1, 
(b) single-quantum coherence of spin 2, (c)
zero-quantum coherence and 
(d) double-quantum coherence of the BTC acid molecule.
}
\label{btp_decay}
\end{figure}

\begin{figure}
\includegraphics[angle=0,scale=1.0]{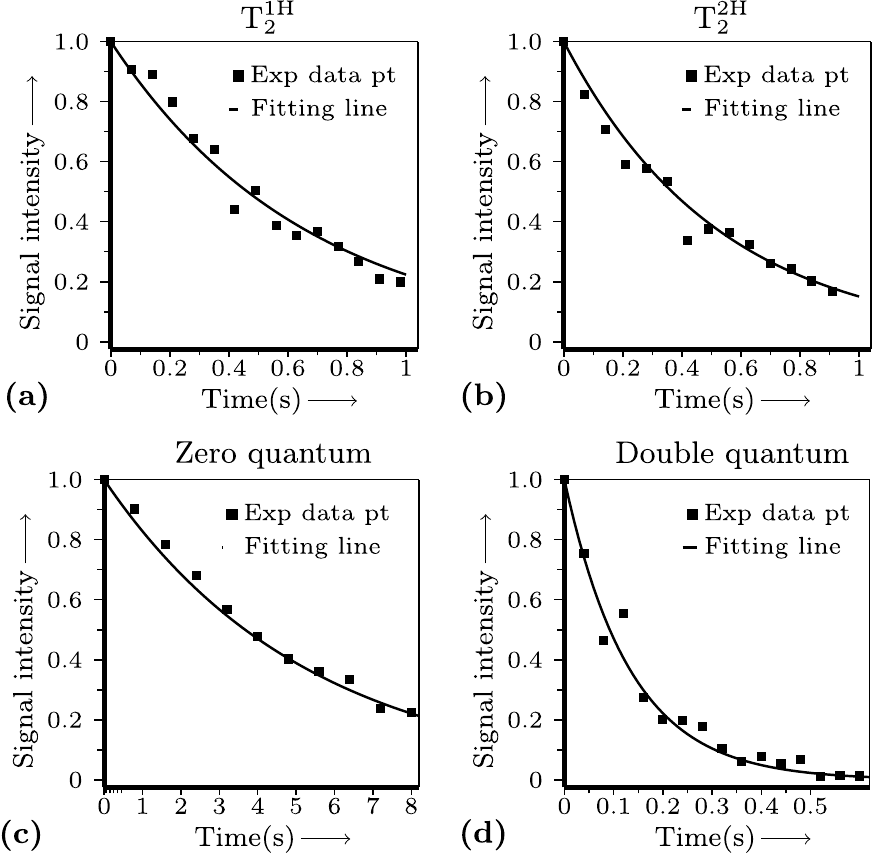}
\caption{Decay of signal intensity with time of the
(a)single-quantum coherence of spin 1, (b) single-quantum
coherence of spin 2, (c)
zero-quantum coherence and (d) double-quantum coherence of the cytosine
molecule.
}
\label{cytosine_decay}
\end{figure}

\begin{table}[h]
\label{gamma-table}
\caption{Correlated phase damping factor present in 
homonuclear two-spin
systems studied, as calculated from fitting the experimental data.}
\vspace*{12pt}
\centering
\renewcommand{\arraystretch}{1.5}
\begin{tabular}{ll}
\hline
Molecule &
Correlated phase damping factor $\gamma_3$ (s$^{-1}$)~~~\\
\hline
BTC acid& $5.876 \pm 1.825$~~~\\
Cytosine & $3.393\pm 1.089$~~~\\
Coumarin&$8.6735 \pm  1.545$~~~\\
\hline
\end{tabular}
\end{table}

\begin{figure}
\includegraphics[angle=0,scale=1.0]{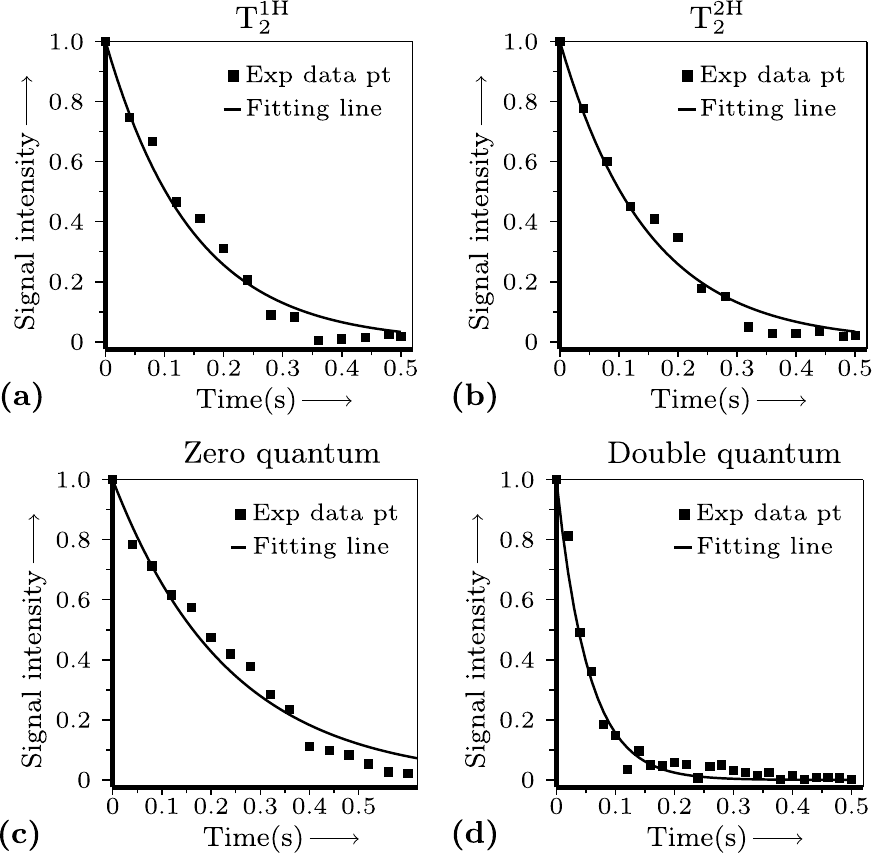}
\caption{Decay of signal intensity with time of the
(a)single-quantum coherence of spin 1, (b) 
single-quantum coherence of spin 2, (c)
zero-quantum coherence and (d) double-quantum coherence of 
the coumarin molecule.
}
\label{coumarine_decay}
\end{figure}

The decay of zero- and double-quantum coherences with time is shown in
Figs.~\ref{btp_decay}-\ref{coumarine_decay}(c)-(d), for the two-spin
homonuclear systems of BTC acid, cytosine and coumarin, respectively.
The experimentally measured values of zero-quantum coherence 
decay rates in these systems
was obtained to be 
$0.430 \pm 0.062$ s$^{-1}$, $0.189\pm 0.004$ s$^{-1}$, and
$4.247\pm 0.267$ s$^{-1}$ 
for the two-spin systems of BTC acid, cytosine, and
coumarin, respectively.
The experimentally measured values of double-quantum coherence 
decay rates in these systems
was obtained to be 
$12.182\pm 1.289$ s$^{-1}$,
$6.975\pm 0.465$ s$^{-1}$, and
$21.594\pm 0.897$ s$^{-1}$, 
for the two-spin systems of BTC acid, cytosine, and
coumarin, respectively.
The correlated phase damping rate $\gamma_3$ 
obtained from fitting the experimental data to
a noise model which incorporates independent and correlated phase
damping as well generalized amplitude damping,
is given in Table~\ref{gamma-table}.
The plots displayed in
Figures~\ref{btp_decay}-\ref{coumarine_decay} show
clear evidence of non-exponential behavior, with 
systematic variations above and below the best fit
exponential. This implies that the Markovian model of noise 
we assumed may not fully capture the noise processes active
in these systems.

For systems of heteronuclear coupled spin-1/2 nuclei, it was previously
shown that the intrinsic NMR noise acting on the spins can be modeled
completely by considering uncorrelated phase damping channels acting
independently on both spins~\cite{childs-pra-01}. 
Our results indicate that this does not
hold true for homonuclear systems of coupled spin-1/2 nuclei, 
where the spins are physically proximate and have identical
gyromagnetic ratios. 
In such cases, the true picture of noise that emerges is a ``correlated''
one, wherein a new phase damping channel acts on both spins together, in
addition to the independent channels acting on each spin separately.
Furthermore, this correlated phase damping channel contributes differentially
to the relaxation rates of the multiple-quantum coherences inherent in the
system. This noise model hence
provides a plausible explanation for why the double-quantum
coherences in homonuclear spin systems decay much faster than the zero-quantum
coherences. On the other hand, heteronuclear spin systems
do not exhibit such effects, indicating that such systems do not have
appreciable correlated phase noise.
\section{Conclusion}
\label{concl}
We used a previously designed
model by Childs {\em et.~al.}~\cite{childs-pra-01} for intrinsic NMR noise in homonuclear two-spin systems as arising from
a correlated phase damping channel acting on both the spins and a generalized
amplitude damping channel acting independently on each spin.  Our results
suggest that the major contribution to spin relaxation in coupled homonuclear
two-spin systems comes from correlated phase damping noise.  The theoretical
model used to describe multiple quantum relaxation in homonuclear two-spin
systems is in good agreement with our experimental data.  We conjecture that
the correlated phase damping behavior exhibited by the multiple quantum
coherences has its origins in contributions from 
dipolar auto-correlated
relaxation as well as from ``remote'' 
cross-correlated interference terms between
two different CSA relaxation mechanism that are present in such systems.
Our results have potential applications to NMR 
relaxation dispersion experiments in large proteins 
and to recent quantum information processing studies which
utilize out-of-time-order (OTOC) correlators, 
where
multiple-quantum coherences play a key
role.
The more general theories
of master equations which are used to describe decoherence processes in open
quantum systems can provide deeper insights into the mechanisms which govern
the relaxation of NMR multi spin relaxation.  The validity of the correlated
phase damping noise model is by no means limited to the two-spin case, but can
be extended to higher-order multiple quanta as well as to larger networks of
magnetically equivalent homonuclear spins.

\begin{acknowledgments}
All experiments were performed on a Bruker
Avance-III 600 MHz FT-NMR spectrometer at the NMR
Research Facility at IISER Mohali.  
\end{acknowledgments}

%
\end{document}